\documentclass[letterpaper]{article} 
\usepackage[preprint]{aaai2027}
\usepackage[hyphens]{url}  
\usepackage{graphicx} 
\usepackage{natbib}  
\usepackage{caption} 
\usepackage{amsmath}
\usepackage{amssymb}
\usepackage{xspace}
\usepackage{algorithm}
\usepackage{algorithmic}

\usepackage{newfloat}
\usepackage{listings}
\DeclareCaptionStyle{ruled}{labelfont=normalfont,labelsep=colon,strut=off} 
\floatstyle{ruled}
\newfloat{listing}{tb}{lst}{}
\floatname{listing}{Listing}

\usepackage{booktabs}

\title{DiffSafeMerge: Mitigating Backdoor Inheritance in Diffusion Model Merging}
\author{
    Jiayang Zhang\textsuperscript{\rm 1}\thanks{Email: jiayang.zhang@std.uestc.edu.cn},
    Ji Guo\textsuperscript{\rm 1},
    Jiachen Li\textsuperscript{\rm 2},
    Wenshu Fan\textsuperscript{\rm 1},
    Wenbo Jiang\textsuperscript{\rm 1}\thanks{*Corresponding author. Email: wenbo\_jiang@uestc.edu.cn}
}
\affiliations{
    \textsuperscript{\rm 1}University of Electronic Science and Technology of China, Chengdu, China\\
    \textsuperscript{\rm 2}Wuhan University of Technology, Wuhan, China
}

\newcommand{\method}{\textsc{DSM}\xspace}
\begin{document}

\maketitle

\begin{abstract}
Unconditional diffusion checkpoint merging assumes benign sources, yet a compromised public checkpoint can transfer a dormant backdoor while clean generation appears normal. Mitigation is difficult without knowing the compromised source, trigger, or target, and broad sanitization may degrade image quality. We introduce DiffSafeMerge (DSM), which uses a small unlabeled clean set and fixed, attack-agnostic stress probes to score source blocks, shrink suspicious contributions toward a trusted reference, and select attenuation under a clean denoising-loss budget. We evaluate four attacks, two datasets, and 21 target conditions. Intended merging already has zero worst-target ASR in 10 of 14 source cases; DSM preserves these outcomes and records no target match in the remaining four over three seeds, including three with baseline ASR of 48--100\%. Among methods with zero worst-target ASR on both datasets, DSM obtains the lowest case-averaged FID in the matched seed-0 comparison.
\end{abstract}


\section{Introduction}

\begin{figure}[t]
    \centering
    \includegraphics[width=\columnwidth]{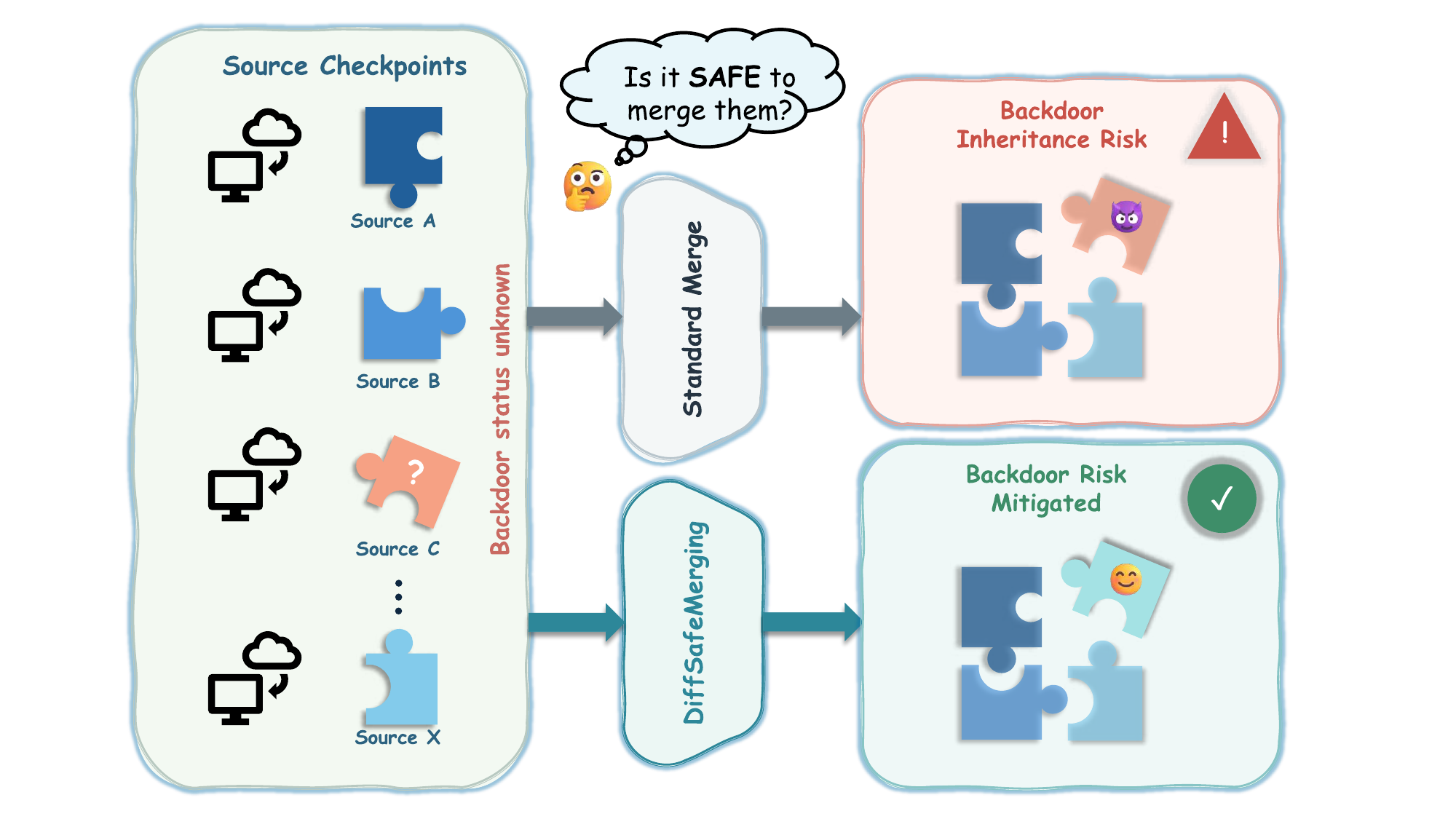}
    \caption{Motivating setting. The defender does not know whether an unverified source is compromised. Standard merging can inherit its hidden backdoor, whereas \method scores and attenuates risky source contributions before merging.}
    \label{fig:scenario}
\end{figure}

Parameter-space merging combines compatible checkpoints without retraining~\cite{wortsman2022model,ilharco2022editing,yadav2023ties}. Public tools make this a repository workflow: Hugging Face documents checkpoint merging and supports adapter merging, including text-to-image LoRAs~\cite{huggingface2024peftmerging}, while Civitai hosts community-shared Stable Diffusion checkpoints and derivatives~\cite{civitai2023platform}. Utility-oriented merging, however, does not account for compromised sources. Diffusion backdoors remain dormant during clean generation and activate from a trigger; BadMerging is designed to survive merging~\cite{chou2023backdoor,chen2023trojdiff,chou2023villandiffusion,han2025uibdiffusion,zhang2024badmerging}. A compromised source may therefore transfer a backdoor despite normal clean samples, although averaging dilutes some attacks. Standard merging cannot predict either outcome (Figure~\ref{fig:scenario}).

At merge time, the defender has the sources, intended coefficients, and a small clean set, but not the compromised source, trigger, or target. Risk may vary across blocks, timesteps, and merge ratios. Existing diffusion defenses detect anomalies, reconstruct triggers, prune a suspected network, or fine-tune it on clean data~\cite{an2024elijah,hao2024diff,mo2024terd,truong2025purediffusion}; they generally start from a suspected checkpoint. Applying them broadly may remove useful denoising parameters, while weight redistribution may amplify another source.

Safety-aware merging in discriminative and language models uses prediction masks, synthetic safety data, alignment vectors, or task-neuron attribution~\cite{yang2025dam,hammoud2024safety,thakkar2025mergealign,ma2025led,djuhera2026safemerge}. DAM relies on classification responses~\cite{yang2025dam}, which unconditional diffusion lacks; source effects also vary across timesteps and merge ratios. Source assessment must therefore avoid attack-specific supervision while controlling distributional change.

We hypothesize that a risky source block produces a stress-amplified response that remains stable as its merge weight changes, whereas removing a useful block incurs a denoising cost. We therefore introduce DiffSafeMerge (DSM). Its Ratio-Scanned Stress-Safe (R3S) score combines source--block stress amplification and directional consistency across fixed probes and ratio paths. Clean utility discounts blocks whose removal harms denoising. DSM then applies a source-level floor, shrinks suspicious contributions toward the trusted reference without weight redistribution, and selects the attenuation strength under a clean-loss budget.

Hyperparameters are fixed after tuning on two BadMerging CIFAR-10 cases. Intended merging has zero worst-target ASR in 10 of 14 source cases and nonzero ASR in four: three at 48--100\% and one at 0.2\%. Across four attacks, two datasets, and 21 target conditions, DSM preserves the former outcomes and records no target match in the latter over three seeds. Among methods with zero worst-target ASR on both datasets, it has the lowest case-averaged FID.

\section{Related Work}

\subsection{Backdoor Attacks and Defenses in Diffusion Models}

Classifier backdoors bind a trigger to an attacker-chosen output while preserving clean accuracy~\cite{gu2017badnets,liu2018trojaning}. Defenses and detectors prune dormant neurons, reconstruct triggers, perturb inputs, stimulate internal units, or adversarially perturb neurons~\cite{liu2018finepruning,wang2019neuralcleanse,gao2019strip,liu2019abs,wu2021adversarial}.

Diffusion attacks transfer this behavior to iterative generation. BadDiffusion and TrojDiff bind triggers to image targets~\cite{chou2023backdoor,chen2023trojdiff}; VillanDiffusion covers several formulations and samplers~\cite{chou2023villandiffusion}; and UIBDiffusion studies imperceptible triggers~\cite{han2025uibdiffusion}. BadMerging further optimizes an attack to survive merging~\cite{zhang2024badmerging}. Existing defenses analyze timestep-wise shifts, invert triggers, or purify a suspected checkpoint~\cite{an2024elijah,hao2024diff,mo2024terd,truong2025purediffusion}. They do not attribute risk across unverified sources.

\subsection{Model Merging and Safety-Aware Merging}

Utility-oriented merging includes averaging and task arithmetic~\cite{wortsman2022model,ilharco2022editing}, Fisher and regression-based fusion~\cite{matena2022fisher,jin2023dataless}, and conflict-aware or adaptive schemes~\cite{yadav2023ties,yang2024adamerging,xu2024trainingfree}. DMM distills diffusion teachers into one model~\cite{song2025dmm}; MaxFusion combines aligned features at inference~\cite{nair2024maxfusion}.

Backdoor security during merging includes dilution~\cite{arora2024freelunch}, post-merge emergence~\cite{wang2025purity}, vector subtraction~\cite{pawlak2025backdoor}, and feature-path optimization~\cite{zhu2026lfpm}. DAM is the closest merge-time formulation: it learns a mask from task-specific clean data, classification responses, and synthesized perturbations, then resets selected task-vector parameters~\cite{yang2025dam}. A diffusion adaptation must redefine perturbation placement, timestep-dependent denoising and mask objectives, and source modification; replacing only the loss would not reproduce DAM. We therefore discuss its threat model without reporting an unvalidated numerical redesign.

Language-model merging can propagate misalignment or backdoors~\cite{hammoud2024safety,yuan2025mergehijacking}. Mitigations use synthetic safety data, alignment vectors, task-neuron attribution, or selective layers~\cite{hammoud2024safety,thakkar2025mergealign,ma2025led,djuhera2026safemerge}, all language-specific signals. DSM uses diffusion denoising responses.

\section{Preliminaries}
\label{sec:problem}

\subsection{Diffusion Model Merging}

Let $\theta_0$ be a trusted common reference and $\{\theta_i\}_{i=1}^{M}$ be architecture-compatible unconditional diffusion checkpoints derived from it. Each source defines a task vector $\Delta_i=\theta_i-\theta_0$. Given intended weights $\{\alpha_i\}_{i=1}^{M}$, the merge without a safety intervention is
\begin{equation}
\theta_{\mathrm{int}}=\theta_0+\sum_{i=1}^{M}\alpha_i\Delta_i,
\ \ \ \alpha_i\geq 0,\quad \sum_i\alpha_i=1,
\label{eq:intended}
\end{equation}
where $\alpha_i$ is fixed by the utility objective. This intended merge is the reference point for measuring any safety-related change.

\subsection{Threat Model}

\noindent\textbf{System Model.}
We consider an automated service that combines architecture-compatible diffusion checkpoints from public repositories or multiple contributors and returns one deployable checkpoint. Source admission and license checks occur upstream. The checkpoints share a trusted reference $\theta_0$, stored independently by the curator, and the intended weights $\{\alpha_i\}$ are fixed before defense.

\noindent\textbf{Adversary.}
The adversary may control the training or post-training modification of one or more sources and choose the poisoning procedure, trigger, and target. Its goal is target matching under triggered sampling while ordinary generation remains plausible. It may know $\theta_0$, the merge rule and weights, and the defense; our adaptive test further gives it the R3S objective. It cannot alter the curator's reference, exact clean samples, or merge-time computation. We study parameter backdoors in loadable checkpoints, not malicious serialization or executable payloads.

\noindent\textbf{Defender.}
The defender has white-box access to the sources, reference, intended weights, and a small unlabeled clean set $\mathcal{C}$. It may use automated denoising evaluations and merge-time parameter transformations, subject to four constraints once a merge job is admitted. \emph{No suspicion-only rejection:} source exclusion cannot serve as the defense. \emph{No exhaustive audit:} per-source manual inspection, trigger reconstruction, and attack-family search are outside the service's latency budget. \emph{No source retraining:} submitted checkpoints cannot be retrained or fine-tuned. \emph{Unknown attack:} the compromised subset, poisoning rate, attack family, trigger, and target are unavailable, and all sources may be benign. The defense must return a single checkpoint from the admitted sources.

\noindent\textbf{Security Goal.}
The defense seeks a deterministic merge-time transformation
\begin{equation}
\begin{aligned}
\theta_{\mathrm{safe}}
&=\mathcal{S}(\theta_0,\{\theta_i\},\{\alpha_i\},\mathcal{C}),\\
L_{\mathrm{clean}}(\theta_{\mathrm{safe}})
&\leq (1+\tau)L_{\mathrm{clean}}(\theta_{\mathrm{int}}),
\end{aligned}
\label{eq:safety-objective}
\end{equation}
where $L_{\mathrm{clean}}$ is the denoising loss estimated on $\mathcal{C}$ and $\tau$ is a predeclared utility budget. The security objective is to reduce inherited target-matching behavior without attack-specific supervision while preserving clean utility. This is a mitigation objective rather than a certificate of complete removal.

\section{DiffSafeMerge}
\label{sec:method}

\subsection{Overview}

\begin{figure*}[t]
    \centering
    \includegraphics[width=0.95\textwidth]{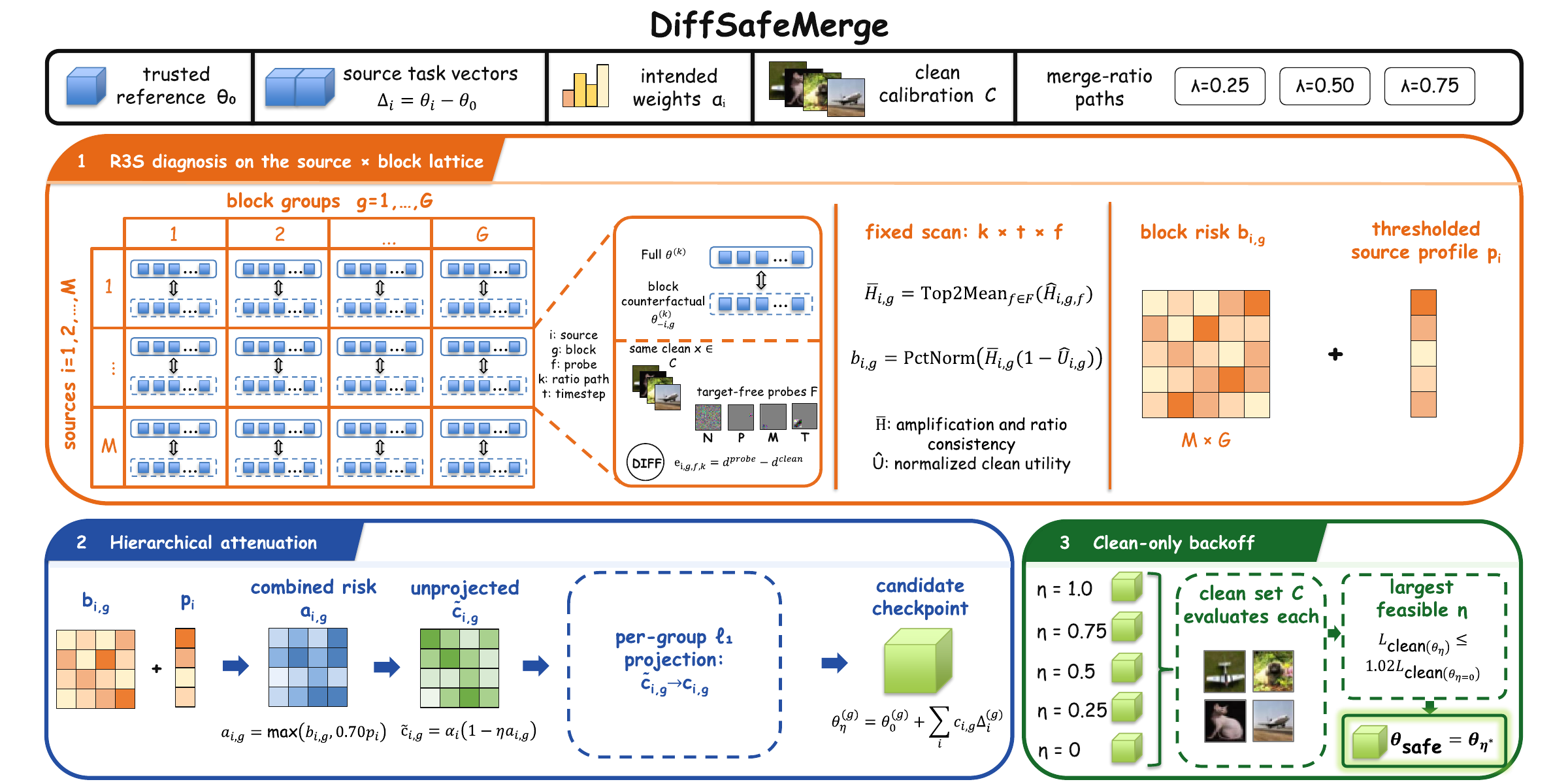}
    \caption{\method pipeline. R3S scores source--block risk from fixed ratio paths and target-free probes. Hierarchical aggregation guides bounded, non-renormalized shrinkage, and clean-only backoff selects the strongest candidate within the denoising-loss budget.}
    \label{fig:overview}
\end{figure*}

\method scores source--block contributions using attack-agnostic response differences while limiting changes to clean denoising. Its working hypothesis is that a risky block produces an amplified response under stress that remains directionally stable as its source weight changes. As Figure~\ref{fig:overview} shows, DSM evaluates source--block counterfactuals across fixed probes, timesteps, and ratio paths, then combines stress amplification, ratio consistency, and clean utility through R3S. Hierarchical aggregation limits source-level masking, while bounded non-renormalized shrinkage and clean-only backoff constrain the intervention.

\begin{algorithm}[t]
\caption{DiffSafeMerge (DSM): Safety-Aware Diffusion Checkpoint Merging}
\label{alg:dsm}
\begin{algorithmic}[1]
\REQUIRE Reference $\theta_0$, sources $\{\theta_i\}$, weights
$\{\alpha_i\}$, clean set $\mathcal{C}$, and candidate strengths
$\mathcal{E}$
\STATE Form task vectors $\Delta_i=\theta_i-\theta_0$
\FOR{each evaluation unit $q$, source $i$, group $g$, and probe $f$}
    \STATE Evaluate matched clean and stressed block counterfactuals
    \STATE Accumulate stress residuals and clean removal costs
\ENDFOR
\STATE Compute amplification $A$, stability $S$, and block risk $b$
\STATE Add the source-level floor to obtain $a_{i,g}$
\FOR{each $\eta\in\mathcal{E}$}
    \STATE Construct $\theta_\eta$ by bounded, non-renormalized shrinkage
    \STATE Evaluate $L_{\mathrm{clean}}(\theta_\eta)$ on $\mathcal{C}$
\ENDFOR
\STATE Select the largest feasible $\eta$ under the clean-loss budget
\RETURN $\theta_{\mathrm{safe}}=\theta_{\eta^*}$
\end{algorithmic}
\end{algorithm}

\subsection{Target-Free Ratio-Scanned Stress Testing}

A checkpoint-level test at the intended ratio may miss weight-dependent responses and does not identify the responsible source block. We therefore remove one block counterfactually and repeat the test along fixed ratio paths. Let $P_g$ select UNet block group $g\in\{1,\ldots,G\}$, $\Delta_i^{(g)}=P_g(\Delta_i)$, and $\boldsymbol{\alpha}^{(k)}$ denote the $k$th scanned weight vector, or ratio path:
\begin{equation}
\begin{aligned}
\theta^{(k)}
&=\theta_0+\sum_{j=1}^{M}\alpha_j^{(k)}\Delta_j,\\
\theta_{-i,g}^{(k)}
&=\theta^{(k)}-\alpha_i^{(k)}P_g(\Delta_i).
\end{aligned}
\label{eq:ratio-counterfactual}
\end{equation}
Here $\sum_j\alpha_j^{(k)}=1$; for two sources, the three paths are $\boldsymbol{\alpha}^{(k)}=(1-\lambda_k,\lambda_k)$ with $\lambda_k\in\{0.25,0.5,0.75\}$. The counterfactual differs only in source $i$'s contribution to group $g$.

For $x_0\in\mathcal{C}$, noise $\epsilon\sim\mathcal{N}(0,I)$, and timestep $t$, we form matched clean and stressed inputs
\begin{equation}
\begin{aligned}
x_t
&=\sqrt{\bar\alpha_t}\,x_0
+\sqrt{1-\bar\alpha_t}\,\epsilon,\\
\widetilde x_{t,f}
&=\sqrt{\bar\alpha_t}\,x_0
+\sqrt{1-\bar\alpha_t}\,\widetilde\epsilon_f .
\end{aligned}
\label{eq:stress-inputs}
\end{equation}
The fixed families $f\in\mathcal{F}$ are noise, patch, mask-overlay, and texture. Their energy is matched across families, and they are constructed without trigger or target information. The exact probe construction, energy level, and timestep choices are given in Supplementary Section~A.2.

Each evaluation unit $q=(B_q,t_q,k_q)$ contains a fixed clean mini-batch
$B_q\subset\mathcal{C}$, timestep $t_q$, and scanned weight vector $k_q$.
Its seeded noise and probes are shared by the full and counterfactual
checkpoints.

Let $\epsilon_\theta(x,t)$ be the predicted noise. We use
\begin{equation}
\begin{aligned}
\ell_{\mathrm{clean}}^{B}(\theta;B,t,\epsilon)
&=\operatorname{MSE}_{B,C,H,W}\!\left(
\epsilon_\theta(x_t,t),\epsilon\right),\\
L_{\mathrm{clean}}(\theta)
&=\mathbb{E}_{B\subset\mathcal{C},\,t\in\mathcal{T},\,\epsilon}
\left[\ell_{\mathrm{clean}}^{B}(\theta;B,t,\epsilon)\right].
\end{aligned}
\label{eq:cleanloss}
\end{equation}
The superscript $B$ makes the mini-batch averaging explicit.

The matched full-versus-counterfactual responses are
\begin{equation}
\begin{aligned}
d^{\mathrm{clean}}_{i,g,q}
&=\epsilon_{\theta^{(k_q)}}(x_{t_q},t_q)
-\epsilon_{\theta_{-i,g}^{(k_q)}}(x_{t_q},t_q),\\
e_{i,g,f,q}
&=\epsilon_{\theta^{(k_q)}}(\widetilde x_{t_q,f},t_q)
-\epsilon_{\theta_{-i,g}^{(k_q)}}(\widetilde x_{t_q,f},t_q)
-d^{\mathrm{clean}}_{i,g,q},\\
\bar e_{i,g,f,k}
&=\frac{1}{|\mathcal{Q}_k|}
\sum_{q\in\mathcal{Q}_k}e_{i,g,f,q},
\quad
\mathcal{Q}_k=\{q\in\mathcal{Q}:k_q=k\}.
\end{aligned}
\label{eq:stress-residual}
\end{equation}
The residual removes the block's clean response, and $\bar e_{i,g,f,k}$ averages evaluation units on the same ratio path. Matched inputs isolate $(i,g)$; scanning timesteps and paths can reveal effects missed at one operating point.

\subsection{Source--Block Risk Scoring with R3S}

Stress sensitivity alone is ambiguous because probe noise and useful blocks can both cause large changes. R3S therefore requires amplification over the clean response and directional consistency across merge ratios:
\begin{equation}
\begin{aligned}
A_{i,g,f}
&=\underset{q\in\mathcal{Q}}{Q^{\mathrm{rank}}_{0.25}}
\left[
\log\!\left(
1+\frac{\operatorname{MSE}(e_{i,g,f,q})}
{\max\{\operatorname{MSE}(d^{\mathrm{clean}}_{i,g,q}),
\varepsilon\}}
\right)
\right],\\
S_{i,g,f}
&=\operatorname{mean}_{k<\ell}
\left[\max\!\left(0,
\cos(\bar e_{i,g,f,k},\bar e_{i,g,f,\ell})
\right)\right],\\
H_{i,g,f}&=A_{i,g,f}S_{i,g,f},
\end{aligned}
\label{eq:r3s}
\end{equation}
where $\varepsilon>0$ stabilizes the ratio and
$Q^{\mathrm{rank}}_{0.25}$ is the rank-based lower quartile defined in
Supplementary Section~A.2. The quartile reduces the influence of isolated spikes,
while the clipped cosine retains aligned responses. After clipped
median--MAD normalization within each family, we aggregate the two largest
family scores:
\begin{equation}
\bar H_{i,g}
=\operatorname{Top2Mean}_{f\in\mathcal{F}}
\left(\widehat H_{i,g,f}\right).
\label{eq:family-aggregate}
\end{equation}

To distinguish suspicious sensitivity from clean utility, we measure the positive loss increase under the same removal:
\begin{equation}
\begin{aligned}
U_{i,g}
&=\frac{1}{|\mathcal{Q}|}\sum_{q\in\mathcal{Q}}\Big[
\ell_{\mathrm{clean}}^{B}(
\theta_{-i,g}^{(k_q)};B_q,t_q,\epsilon_q)\\
&\hspace{5.4em}{}-\ell_{\mathrm{clean}}^{B}(
\theta^{(k_q)};B_q,t_q,\epsilon_q)\Big]_{+},\\
r_{i,g}&=\bar H_{i,g}(1-\widehat U_{i,g}),\\
b_{i,g}&=\operatorname{PctNorm}(r_{i,g})\in[0,1].
\end{aligned}
\label{eq:risk}
\end{equation}
Here $(u)_+=\max(u,0)$, $\widehat U_{i,g}$ is the robustly normalized utility cost, and $\operatorname{PctNorm}$ maps scores from the 50th to the 90th percentile over source--block pairs. The utility factor discounts blocks whose removal harms clean denoising, so high risk requires persistent multi-probe evidence without a comparable clean penalty. This combination favors repeatable stressed responses over isolated probe effects and clean-critical blocks.

\subsection{Hierarchical Risk Aggregation}

A global block ranking can overlook one source when another dominates the score range. We retain block resolution and add a source-level floor. Let $s_i=\max_g r_{i,g}$ and $s_{\max}=\max_j s_j$:
\begin{equation}
\begin{aligned}
p_i&=
\begin{cases}
s_i/s_{\max},
& s_{\max}\geq\varepsilon_s
\ \text{and}\ s_i\geq h s_{\max},\\
0, & \text{otherwise},
\end{cases}\\
a_{i,g}&=\max(b_{i,g},\lambda p_i).
\end{aligned}
\label{eq:hierarchical-risk}
\end{equation}
The profile $p_i\in[0,1]$ is not a probability. The two terms retain block-level scores and assign a floor to sources with sufficient evidence; the threshold leaves all profiles at zero when evidence is negligible.

\subsection{Bounded Non-Renormalized Shrinkage}

Unconstrained scores can move a block too far from the intended merge, while renormalization can transfer removed weight to another unverified source. For attenuation strength $\eta\in[0,1]$, we first shrink each coefficient:
\begin{equation}
\begin{aligned}
\widetilde c_{i,g}&=\alpha_i(1-\eta a_{i,g}),\\
D_g&=\sum_i|\widetilde c_{i,g}-\alpha_i|,\\
\gamma_g&=
\begin{cases}
1, & D_g\leq\delta,\\
\delta/D_g, & D_g>\delta.
\end{cases}
\end{aligned}
\label{eq:projection-scale}
\end{equation}
Here $\delta$ bounds the total coefficient change in each group. The projected coefficients and candidate checkpoint are
\begin{equation}
\begin{aligned}
c_{i,g}
&=\alpha_i+\gamma_g(\widetilde c_{i,g}-\alpha_i),\\
\theta_\eta^{(g)}
&=\theta_0^{(g)}
+\sum_{i=1}^{M}c_{i,g}\Delta_i^{(g)},
\quad g=1,\ldots,G.
\end{aligned}
\label{eq:candidate}
\end{equation}
We do not renormalize $\{c_{i,g}\}_i$: removed contribution falls back to $\theta_0^{(g)}$ instead of increasing another source coefficient. The projection bounds the per-group $\ell_1$ coefficient deviation from the intended merge; fixed normalization rules, hierarchy thresholds, and the projection bound appear in Supplementary Section~A.2.

\subsection{Clean-Only Quality Backoff}

Risk is only a surrogate, so even a well-ranked intervention can damage generation. We construct candidates for $\mathcal{E}=\{1,0.75,0.5,0.25,0\}$ and accept the strongest one within the clean-loss budget. With $L_0=L_{\mathrm{clean}}(\theta_{\eta=0})$ and $\tau=0.02$,
\begin{equation}
\eta^*
=\max\left\{
\eta\in\mathcal{E}:L_{\mathrm{clean}}(\theta_\eta)
\leq(1+\tau)L_0
\right\}.
\label{eq:backoff}
\end{equation}
The output is $\theta_{\mathrm{safe}}=\theta_{\eta^*}$. Selection uses only $\mathcal{C}$, never triggered samples, attack success, or target labels; $\eta=0$ keeps the feasible set nonempty. This check enforces the empirical clean-loss constraint instead of assuming that risk ranking preserves clean quality. The method returns one checkpoint with no additional inference cost. Algorithm~\ref{alg:dsm} summarizes the procedure; fixed settings and cost analysis appear in Supplementary Sections~A.2 and~A.3.

\section{Experiments}
\label{sec:experiments}

\subsection{Experimental Settings}

\paragraph{Datasets and Attack Settings.}
We evaluate unconditional DDPMs~\cite{ho2020denoising} on CIFAR-10~\cite{krizhevsky2009learning} and CelebA-HQ~\cite{karras2018progressive} under BadMerging, BadDiffusion, UIBDiffusion, and VillanDiffusion. In S1, a clean checkpoint is merged with a backdoored checkpoint and evaluated on HAT. S2 merges two backdoored checkpoints and evaluates HAT and CAT. The formal evaluation contains eight CIFAR-10 source cases with 12 target conditions and six CelebA-HQ cases with nine target conditions.

\paragraph{Baselines.}
We compare \method with three merging controls (Diffusion Soup, DMM-style distillation, and MaxFusion-style score fusion) and two sanitization baselines (ANP and clean fine-tuning). Implementations and matching rules are listed in Supplementary Section~B.2.

\paragraph{Evaluation Protocol.}
We tune hyperparameters on the BadMerging CIFAR-10 S1 and S2 development cases and fix them elsewhere. Formal runs merge two sources at $0.5/0.5$ for 1000 steps with seeds 0--2. Supplementary Section~B lists checkpoints, calibration and sample sizes, and the three unavailable UIBDiffusion CelebA-HQ conditions.

\paragraph{Metrics.}
ASR is the fraction of triggered generations with target MSE below $0.05$; we also report target MSE, Structural Similarity Index (SSIM)~\cite{wang2004image}, and clean Fr\'{e}chet Inception Distance (FID)~\cite{heusel2017gans}. \method uses three seeds and two-sided Wilson intervals; baselines use seed 0. Mean and worst ASR aggregate target conditions. FID is averaged by source case, so S2 counts once. Since FID depends on checkpoint, sample count, preprocessing, and sampling, we compare it only within this matched protocol, not with training-time benchmarks, and make no generation-SOTA claim. Selection uses no evaluation metric or triggered sample; Supplementary Section~B.3 gives the protocol.

\begin{table*}[t]
\centering
\small
\setlength{\tabcolsep}{1.5pt}
\begin{tabular}{l*{11}{r}}
\toprule
& \multicolumn{2}{c}{BadMerging} & \multicolumn{2}{c}{BadDiffusion} & \multicolumn{2}{c}{UIBDiffusion} & \multicolumn{2}{c}{VillanDiffusion} & \multicolumn{3}{c}{Across attacks} \\
\cmidrule(lr){2-3}\cmidrule(lr){4-5}\cmidrule(lr){6-7}\cmidrule(lr){8-9}\cmidrule(lr){10-12}
Method & FID$\downarrow$ & ASR$\downarrow$ & FID$\downarrow$ & ASR$\downarrow$ & FID$\downarrow$ & ASR$\downarrow$ & FID$\downarrow$ & ASR$\downarrow$ & Avg. FID & Mean ASR & Worst ASR \\
\midrule
Diffusion Soup & 429.4 & 100.00 & 48.7 & 0.00 & 46.3 & 0.00 & 47.0 & 0.00 & 142.9 & 25.00 & 100.00 \\
DMM & 62.0 & 99.51 & 51.4 & 0.00 & 47.4 & 0.00 & 45.7 & 0.00 & 51.6 & 24.88 & 99.51 \\
MaxFusion & 56.0 & 0.00 & 51.9 & 0.00 & 51.5 & 0.00 & 51.6 & 0.00 & 52.8 & 0.00 & 0.00 \\
\midrule
Clean fine-tuning & 53.7 & 0.20 & 56.1 & 0.00 & 54.7 & 0.00 & 57.3 & 0.00 & 55.4 & 0.05 & 0.20 \\
ANP & 247.5 & 0.00 & 213.8 & 0.00 & 268.7 & 0.00 & 231.7 & 0.00 & 240.4 & 0.00 & 0.00 \\
\midrule
\method & 50.1 & 0.00 & 51.5 & 0.00 & 49.7 & 0.00 & 47.1 & 0.00 & 49.6 & 0.00 & 0.00 \\
\bottomrule

\end{tabular}
\caption{Matched seed-0 CIFAR-10 S1 comparison. ASR is evaluated on HAT; summary columns aggregate the four attacks.}
\label{tab:cifar-s1-comparison}
\end{table*}

\begin{table*}[t]
\centering
\small
\setlength{\tabcolsep}{1.25pt}
\begin{tabular}{ll*{11}{r}}
\toprule
& & \multicolumn{2}{c}{BadMerging} & \multicolumn{2}{c}{BadDiffusion} & \multicolumn{2}{c}{UIBDiffusion} & \multicolumn{2}{c}{VillanDiffusion} & \multicolumn{3}{c}{Across attacks} \\
\cmidrule(lr){3-4}\cmidrule(lr){5-6}\cmidrule(lr){7-8}\cmidrule(lr){9-10}\cmidrule(lr){11-13}
Method & Tgt. & FID$\downarrow$ & ASR$\downarrow$ & FID$\downarrow$ & ASR$\downarrow$ & FID$\downarrow$ & ASR$\downarrow$ & FID$\downarrow$ & ASR$\downarrow$ & Avg. FID & Mean ASR & Worst ASR \\
\midrule
Diffusion Soup & HAT & 433.9 & 100.00 & 48.6 & 0.00 & 45.9 & 0.00 & 49.6 & 0.20 & 144.5 & 25.05 & 100.00 \\
 & CAT & 434.6 & 0.00 & 48.6 & 0.00 & 45.6 & 0.00 & 48.2 & 0.00 & 144.2 & 0.00 & 0.00 \\
\addlinespace[0.1em]
DMM & HAT & 61.5 & 99.02 & 51.1 & 0.00 & 47.0 & 0.00 & 46.7 & 0.00 & 51.6 & 24.76 & 99.02 \\
 & CAT & 62.2 & 0.00 & 52.7 & 0.00 & 47.0 & 0.00 & 46.7 & 0.00 & 52.2 & 0.00 & 0.00 \\
\addlinespace[0.1em]
MaxFusion & HAT & 62.6 & 99.61 & 52.7 & 98.05 & 49.6 & 93.07 & 49.3 & 100.00 & 53.6 & 97.68 & 100.00 \\
 & CAT & 62.8 & 0.00 & 53.3 & 0.00 & 49.6 & 0.00 & 49.1 & 0.00 & 53.7 & 0.00 & 0.00 \\
\midrule
Clean fine-tuning & HAT & 54.7 & 92.19 & 54.0 & 0.00 & 56.2 & 0.00 & 61.9 & 0.00 & 56.7 & 23.05 & 92.19 \\
 & CAT & 53.8 & 0.00 & 54.3 & 0.00 & 57.4 & 0.00 & 62.2 & 0.00 & 56.9 & 0.00 & 0.00 \\
\addlinespace[0.1em]
ANP & HAT & 239.2 & 0.00 & 230.3 & 0.00 & 219.6 & 0.00 & 217.0 & 0.00 & 226.5 & 0.00 & 0.00 \\
 & CAT & 248.7 & 0.00 & 303.2 & 0.00 & 217.6 & 0.00 & 295.6 & 0.00 & 266.3 & 0.00 & 0.00 \\
\midrule
\method & HAT & 59.8 & 0.00 & 50.0 & 0.00 & 46.6 & 0.00 & 46.9 & 0.00 & 50.8 & 0.00 & 0.00 \\
 & CAT & 59.8 & 0.00 & 50.0 & 0.00 & 46.6 & 0.00 & 46.9 & 0.00 & 50.8 & 0.00 & 0.00 \\
\bottomrule

\end{tabular}
\caption{Matched seed-0 CIFAR-10 S2 comparison. HAT and CAT are reported separately; target-similarity results appear in Supplementary Section~C.2.}
\label{tab:cifar-s2-comparison}
\end{table*}

\subsection{Main Results}

\paragraph{Overall Safety and Utility.}
Intended merging has zero worst-target ASR in 10 of 14 source cases, but retains 48--100\% in three BadMerging cases and 0.2\% in VillanDiffusion CIFAR-10 S2. DSM preserves the first 10 outcomes and records no target match in the remaining four over three seeds. The zero-success counts are 0/3,072 per CIFAR-10 condition and 0/768 per CelebA-HQ condition, with 95\% Wilson upper bounds of 0.125\% and 0.498\%. Case-averaged FID is 49.67 and 164.07, respectively. Supplementary Section~C.1 gives all conditions by seed.

Clean quality is less uniform: CIFAR-10 FID ranges from 46.38 to 57.11, while the two CelebA-HQ VillanDiffusion cases reach 339.97 and 186.00. ASR measures target matching; FID measures standard no-trigger generation.

\begin{figure}[!ht]
\centering
\includegraphics[width=\columnwidth]{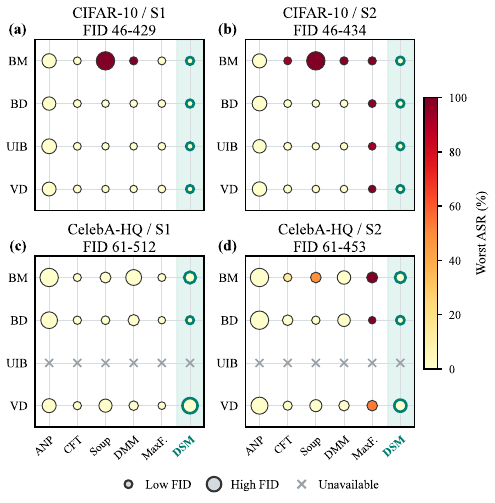}
\caption{Matched seed-0 comparison. Color denotes worst-target ASR and marker area panel-scaled FID; crosses mark unavailable conditions. CFT and MaxF. denote clean fine-tuning and MaxFusion. Tables~\ref{tab:cifar-s1-comparison}--\ref{tab:celeba-s2-comparison} give exact values.}
\label{fig:matched-baseline-matrix}
\end{figure}

\begin{figure}[!ht]
\centering
\includegraphics[width=\columnwidth]{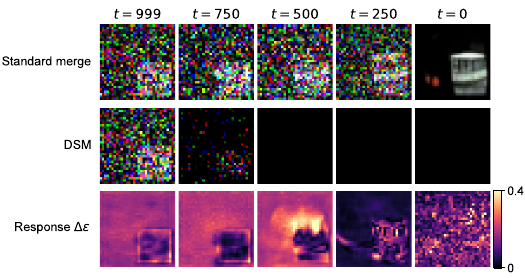}
\caption{Selected successful Diffusion Soup trajectory from held-out VillanDiffusion CIFAR-10 S2 (BOX\_14, seed 0; case HAT ASR 0.20\%). Under shared noise, DSM departs from this target-reaching example. RMS maps diagnose this trajectory, not event frequency or attention.}
\label{fig:denoising-response}
\end{figure}

\paragraph{Comparison with Merging and Defense Baselines.}
Figure~\ref{fig:matched-baseline-matrix} and Tables~\ref{tab:cifar-s1-comparison}--\ref{tab:celeba-s2-comparison} cover 126 method--target evaluations. Only ANP and \method have zero worst-target ASR on both datasets; DSM has lower case-averaged FID: 50.2 versus 243.4 on CIFAR-10 and 166.2 versus 418.0 on CelebA-HQ. MaxFusion has lower FID in all three CelebA-HQ S1 cases, but fuses sources at inference rather than producing one checkpoint and retains target matches in S2. Thus, DSM does not claim per-case FID dominance. DMM reaches zero worst-target ASR on CelebA-HQ at FID 207.8, but not on both datasets. Supplementary Section~C.2 gives per-case and MSE/SSIM results.

\begin{table*}[!t]
\centering
\small
\setlength{\tabcolsep}{1.45pt}
\begin{tabular}{l*{9}{r}}
\toprule
& \multicolumn{2}{c}{BadMerging} & \multicolumn{2}{c}{BadDiffusion} & \multicolumn{2}{c}{VillanDiffusion} & \multicolumn{3}{c}{Across attacks} \\
\cmidrule(lr){2-3}\cmidrule(lr){4-5}\cmidrule(lr){6-7}\cmidrule(lr){8-10}
Method & FID$\downarrow$ & ASR$\downarrow$ & FID$\downarrow$ & ASR$\downarrow$ & FID$\downarrow$ & ASR$\downarrow$ & Avg. FID & Mean ASR & Worst ASR \\
\midrule
Diffusion Soup & 165.7 & 0.00 & 93.2 & 0.00 & 248.4 & 0.00 & 169.1 & 0.00 & 0.00 \\
DMM & 374.1 & 0.00 & 176.7 & 0.00 & 117.7 & 0.00 & 222.8 & 0.00 & 0.00 \\
MaxFusion & 73.4 & 0.00 & 60.6 & 0.00 & 66.9 & 0.00 & 67.0 & 0.00 & 0.00 \\
\midrule
Clean fine-tuning & 76.0 & 0.00 & 83.1 & 0.00 & 81.8 & 0.00 & 80.3 & 0.00 & 0.00 \\
ANP & 512.5 & 0.00 & 418.6 & 0.00 & 284.9 & 0.00 & 405.3 & 0.00 & 0.00 \\
\midrule
\method & 187.9 & 0.00 & 63.8 & 0.00 & 339.8 & 0.00 & 197.2 & 0.00 & 0.00 \\
\bottomrule

\end{tabular}
\caption{Matched seed-0 CelebA-HQ S1 over three available attacks; ASR is evaluated on HAT and UIBDiffusion is unavailable.}
\label{tab:celeba-s1-comparison}

\centering
\small
\setlength{\tabcolsep}{1.25pt}
\begin{tabular}{ll*{9}{r}}
\toprule
& & \multicolumn{2}{c}{BadMerging} & \multicolumn{2}{c}{BadDiffusion} & \multicolumn{2}{c}{VillanDiffusion} & \multicolumn{3}{c}{Across attacks} \\
\cmidrule(lr){3-4}\cmidrule(lr){5-6}\cmidrule(lr){7-8}\cmidrule(lr){9-11}
Method & Tgt. & FID$\downarrow$ & ASR$\downarrow$ & FID$\downarrow$ & ASR$\downarrow$ & FID$\downarrow$ & ASR$\downarrow$ & Avg. FID & Mean ASR & Worst ASR \\
\midrule
Diffusion Soup & HAT & 140.0 & 48.05 & 79.1 & 0.00 & 186.8 & 0.00 & 135.3 & 16.02 & 48.05 \\
 & CAT & 142.9 & 0.00 & 85.3 & 0.00 & 185.2 & 0.00 & 137.8 & 0.00 & 0.00 \\
\addlinespace[0.1em]
DMM & HAT & 236.7 & 0.00 & 196.9 & 0.00 & 137.8 & 0.00 & 190.5 & 0.00 & 0.00 \\
 & CAT & 233.5 & 0.00 & 213.0 & 0.00 & 138.3 & 0.00 & 194.9 & 0.00 & 0.00 \\
\addlinespace[0.1em]
MaxFusion & HAT & 154.1 & 100.00 & 61.4 & 0.00 & 140.5 & 0.00 & 118.6 & 33.33 & 100.00 \\
 & CAT & 155.0 & 0.00 & 60.5 & 100.00 & 138.7 & 53.52 & 118.1 & 51.17 & 100.00 \\
\midrule
Clean fine-tuning & HAT & 91.9 & 14.06 & 124.1 & 0.00 & 98.0 & 0.00 & 104.7 & 4.69 & 14.06 \\
 & CAT & 70.0 & 0.00 & 157.5 & 0.00 & 106.8 & 0.00 & 111.4 & 0.00 & 0.00 \\
\addlinespace[0.1em]
ANP & HAT & 416.3 & 0.00 & 514.1 & 0.00 & 391.3 & 0.00 & 440.6 & 0.00 & 0.00 \\
 & CAT & 489.6 & 0.00 & 391.2 & 0.00 & 381.4 & 0.00 & 420.7 & 0.00 & 0.00 \\
\midrule
\method & HAT & 141.7 & 0.00 & 77.6 & 0.00 & 186.6 & 0.00 & 135.3 & 0.00 & 0.00 \\
 & CAT & 141.7 & 0.00 & 77.6 & 0.00 & 186.6 & 0.00 & 135.3 & 0.00 & 0.00 \\
\bottomrule

\end{tabular}
\caption{Matched seed-0 CelebA-HQ S2 over three available attacks; HAT and CAT are separate and UIBDiffusion is unavailable.}
\label{tab:celeba-s2-comparison}

\includegraphics[width=0.76\textwidth]{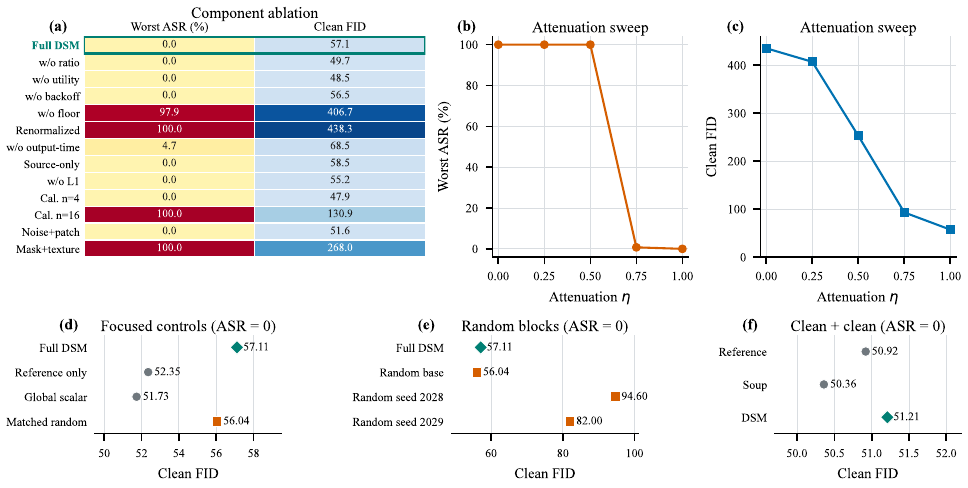}
\captionof{figure}{Development-case diagnostics on BadMerging CIFAR-10 S2: (a) components; (b,c) attenuation; (d) focused controls; (e) random-block stability; and (f) clean--clean merging. Three seeds; complete numerical results appear in Supplementary Section~D.1.}
\label{fig:component-analysis}
\end{table*}

\subsection{More Analysis}

\paragraph{Denoising Response.}
Figure~\ref{fig:denoising-response} shows a selected successful Diffusion Soup trajectory and DSM under shared noise. It is a mechanistic example, not a frequency estimate; the RMS maps are neither attention nor causal attribution. Supplementary Section~D.1 decomposes R3S.

\paragraph{Component Analysis.}
On this development case, removing the source floor, renormalizing source weights, and excluding output-time evidence yield HAT ASRs of 97.95\%, 100\%, and 4.66\% (Figure~\ref{fig:component-analysis}). These are case-specific sensitivities. Sixteen calibration images or mask--texture probes give 100\% worst ASR; four images, noise--patch probes, and the ratio-stability and backoff ablations remain at zero. Supplementary Section~D.1 gives complete results.

\paragraph{Sensitivity and Adaptive Stress Test.}
On this case, ASR is 100\% through $\eta=0.5$, 0.749\% at $\eta=0.75$, and zero at $\eta=1$ as FID falls from 435.41 to 57.11 (Figure~\ref{fig:component-analysis}b--c). Full DSM, reference-only, scalar attenuation, and matched-random all reach zero ASR, with FIDs of 57.11, 52.35, 51.73, and 56.04. Reference-only discards source updates and is not a capability-preserving merge. Across three matched-random choices, zero ASR accompanies FIDs of 56.04--94.60, supporting only unstable quality rather than uniqueness or necessity of R3S. Clean--clean reference, soup, and DSM yield FIDs of 50.92, 50.36, and 51.21 with 100\% valid clean outputs; an R3S-aware test retains $0.0\pm0.0\%$ ASR as FID changes from $49.58\pm0.54$ to $52.47\pm0.47$.

\paragraph{Clean Utility and Efficiency.}

Clean utility uses no-trigger FID, validity, and denoising loss; security uses triggered target matching. All 25 audited clean configurations are valid. DSM adds no inference pass; costs appear in Supplementary Section~D.

\section{Conclusion}

DiffSafeMerge mitigates unknown source contamination with ratio-scanned scoring and clean-constrained shrinkage. Across four attacks, two datasets, and 21 target conditions, DSM preserves already-diluted zero-ASR cases and records no target match in four inherited-risk cases over three seeds; three have 48--100\% intended-merge ASR. Among methods reaching zero worst-target ASR on both datasets, DSM has the lowest matched seed-0 case-averaged FID.

\bibliography{references}

\clearpage
\appendix
\appendix
\setcounter{secnumdepth}{2}
\section{Additional DiffSafeMerge Details}
\label{app:method-details}

\subsection{Full Procedure}
\label{app:procedure}

\noindent\begin{minipage}{\columnwidth}
\hrule
\vspace{0.35em}
\noindent\textbf{DiffSafeMerge pseudocode}
\vspace{0.2em}
\begin{algorithmic}[1]
\REQUIRE Trusted reference $\theta_0$; source checkpoints
$\{\theta_i\}_{i=1}^{M}$; intended weights $\{\alpha_i\}$;
clean set $\mathcal{C}$; ratio paths, timesteps, probes, and
candidate strengths $\mathcal{E}$
\STATE Form task vectors $\Delta_i=\theta_i-\theta_0$
\FOR{each evaluation unit $q$, source $i$, group $g$, and probe $f$}
    \STATE Evaluate the full and block-counterfactual checkpoints on
    matched clean and stressed inputs
    \STATE Accumulate the clean response, stress residual, and clean
    removal cost
\ENDFOR
\STATE Compute $A_{i,g,f}$, $S_{i,g,f}$, and utility-adjusted block
risk $b_{i,g}$
\STATE Combine block risk with the source profile to obtain
$a_{i,g}$
\FOR{each attenuation strength $\eta\in\mathcal{E}$}
    \STATE Shrink and project coefficients without renormalization
    to construct $\theta_\eta$
    \STATE Evaluate $L_{\mathrm{clean}}(\theta_\eta)$ on
    $\mathcal{C}$
\ENDFOR
\STATE Select the largest feasible $\eta$ under
$L_{\mathrm{clean}}(\theta_\eta)\leq(1+\tau)L_0$
\RETURN $\theta_{\mathrm{safe}}=\theta_{\eta^*}$
\end{algorithmic}
\vspace{0.25em}
\hrule
\end{minipage}

\subsection{Fixed Diagnostic and Shrinkage Settings}
\label{app:fixed-settings}

We scan $\lambda\in\{0.25,0.5,0.75\}$, which gives three
weight-vector paths in the two-source experiments. With more than two
sources, each value of $\lambda$ gives one anchored path per source.
We use the discrete diffusion
timesteps nearest to $\{0.25,0.5,0.75\}$ of the training horizon.
The noise, patch, mask-overlay, and texture probes are energy matched
by
$\lVert\widetilde\epsilon_f-\epsilon\rVert_2
=0.1\lVert\epsilon\rVert_2$.
All probes are fixed before evaluation.

For R3S, the numerical stabilizer is $\varepsilon=10^{-12}$.
For sorted values $v_{(1)}\leq\cdots\leq v_{(n)}$, the rank
aggregator in main-paper Eq.~(7) is
\begin{equation}
Q^{\mathrm{rank}}_{0.25}(v)=
\begin{cases}
v_{(2)}, & n=3,\\
v_{(1+\lceil 0.25(n-1)\rceil)}, & \text{otherwise}.
\end{cases}
\label{eq:rank-quantile}
\end{equation}

Both $H_{i,g,f}$ and $U_{i,g}$ use the following robust
normalization. For a vector $v$, let
$m=\operatorname{median}(v)$,
$d=\operatorname{median}(|v-m|)$, $s=\max(v)-m$, and
$q_{\max}=5$. When $d\geq\varepsilon$, we use
\begin{equation}
\operatorname{RobustNorm}(v_j)
=\frac{1}{q_{\max}}\operatorname{clip}\!\left(
\frac{v_j-m}{d+\varepsilon},0,q_{\max}\right).
\label{eq:robust-normalization}
\end{equation}
If $d<\varepsilon$ but $s\geq\varepsilon$, the map instead uses
$\operatorname{clip}((v_j-m)/(s+\varepsilon),0,1)$. It returns an
all-zero vector when both $d$ and $s$ are below $\varepsilon$.
The map is applied within each probe family for $H_{i,g,f}$ and
over all source--block pairs for $U_{i,g}$. For
$p_{50}=\operatorname{quantile}_{0.50}(r)$ and
$p_{90}=\operatorname{quantile}_{0.90}(r)$, if
$p_{90}-p_{50}\geq\varepsilon$ we use
\begin{equation}
\operatorname{PctNorm}(r_j)=
\operatorname{clip}\!\left(
\frac{r_j-p_{50}}{p_{90}-p_{50}+\varepsilon},0,1\right).
\label{eq:percentile-normalization}
\end{equation}
The map returns an all-zero vector when the percentile spread is
below $\varepsilon$.

Hierarchical aggregation uses
$h=0.45$, $\varepsilon_s=10^{-6}$, and $\lambda=0.70$.
The per-group projection bound is $\delta=0.85$.
Clean-only backoff uses
$\mathcal{E}=\{1,0.75,0.5,0.25,0\}$ and $\tau=0.02$.
These values are selected on the development cases described in the
experimental setup and then held fixed.

\subsection{Computational Cost}
\label{app:complexity}

For $M$ sources, $G$ groups, $K$ scanned weight vectors (ratio
paths), $F$ probe families, $T$ timesteps, and $N_B$ clean
mini-batches, a direct
implementation uses
$O(KN_BT(F+1)(1+MG))$ batched UNet forward evaluations.
Candidate construction is linear in the parameter count and $M$.
\method trains neither an attack nor a source model and returns one
checkpoint, so it requires no inference-time ensemble.

\section{Extended Experimental Setup}
\label{app:experimental-setup}

\subsection{Datasets, Checkpoints, and Merge Cases}
\label{app:source-cases}

All evaluated checkpoints are unconditional DDPMs with compatible
architectures and a common trusted reference. CIFAR-10 contains S1 and
S2 cases for BadMerging, BadDiffusion, UIBDiffusion, and
VillanDiffusion. CelebA-HQ contains S1 and S2 cases for BadMerging,
BadDiffusion, and VillanDiffusion. UIBDiffusion CelebA-HQ checkpoints
were unavailable, leaving 14 of 16 source cases and 21 of 24 target
conditions. S1 combines a clean source with one backdoored source and
uses HAT. S2 combines HAT- and CAT-backdoored sources and evaluates
both targets.

The clean calibration set contains eight CIFAR-10 images or four
CelebA-HQ images. It is used for R3S diagnostics and clean-only
backoff, but not for attack reconstruction or target selection. All
formal merges use intended weights $0.5/0.5$. The evaluation uses
seeds 0, 1, and 2, 1000 denoising steps, and 1024 CIFAR-10 or 256
CelebA-HQ samples per seed.

\subsection{Baseline Implementations and Matching}
\label{app:baseline-details}

Diffusion Soup is the intended parameter average. The DMM-style
control initializes a student from this average and distills the
weighted source predictions. The MaxFusion-style control combines
source scores with a variance-dependent gate at inference. ANP
identifies and prunes neurons associated with backdoor behavior.
Clean fine-tuning updates the averaged checkpoint using clean data.

A baseline row enters the comparison only when the source
checkpoints, $0.5/0.5$ weights, evaluation seed, sample count, and
1000-step sampler agree with the \method row. The core matched archive
contains 13 target conditions, and exact seed-0 extensions add the
remaining eight. For saved CelebA-HQ extension runs with more than 256
images, metrics are recomputed on one fixed deterministic 256-image
subset; the original generations are unchanged. The complete
comparison therefore covers 14 source cases and 21 target conditions.
DMM and MaxFusion are implementation-level controls rather than exact
reproductions of every training choice in their original settings.

\subsection{Metrics and Statistical Protocol}
\label{app:metric-protocol}

For a target image $y$, a triggered output $\hat{x}$ is successful
when $\operatorname{MSE}(\hat{x},y)<0.05$. ASR is the empirical
success fraction. We report the mean and standard deviation over
three seeds and a two-sided Wilson interval from the pooled success
count. Target MSE and SSIM describe similarity to the attack target;
they are not measures of general perceptual quality. FID is computed
from clean samples. Dataset-level FID first averages targets within a
source case and then macro-averages cases.

Clean utility is evaluated without a trigger using FID, clean validity,
and clean denoising loss. Clean validity is reported only for standard
no-trigger samples.

\section{Complete Results}
\label{app:complete-results}

\subsection{Per-Seed Main Results}
\label{app:complete-main-results}

Tables~\ref{tab:complete-main-results-cifar} and
\ref{tab:complete-main-results-celeba} separate the three seeds behind
the aggregate main results. In S2, HAT and CAT share the same clean
sample set, so their FID values coincide within a seed; their target
MSE and SSIM remain target-specific.

\begin{table*}[t]
\centering
\small
\setlength{\tabcolsep}{3.8pt}
\begin{tabular}{lllrrrrr}
\toprule
Attack & Scen. & Target & Seed & ASR (\%) & FID & MSE & SSIM \\
\midrule
BadMerging & S1 & HAT & 0 & 0.00 & 50.09 & 0.2406 & 0.0004 \\
BadMerging & S1 & HAT & 1 & 0.00 & 49.63 & 0.2406 & 0.0004 \\
BadMerging & S1 & HAT & 2 & 0.00 & 49.02 & 0.2406 & 0.0004 \\
BadMerging & S2 & HAT & 0 & 0.00 & 59.77 & 0.2406 & 0.0004 \\
BadMerging & S2 & HAT & 1 & 0.00 & 54.71 & 0.2406 & 0.0004 \\
BadMerging & S2 & HAT & 2 & 0.00 & 56.85 & 0.2406 & 0.0004 \\
BadMerging & S2 & CAT & 0 & 0.00 & 59.77 & 0.3611 & 0.0002 \\
BadMerging & S2 & CAT & 1 & 0.00 & 54.71 & 0.3611 & 0.0002 \\
BadMerging & S2 & CAT & 2 & 0.00 & 56.85 & 0.3611 & 0.0002 \\
BadDiffusion & S1 & HAT & 0 & 0.00 & 51.47 & 0.2406 & 0.0004 \\
BadDiffusion & S1 & HAT & 1 & 0.00 & 50.77 & 0.2406 & 0.0004 \\
BadDiffusion & S1 & HAT & 2 & 0.00 & 50.70 & 0.2406 & 0.0004 \\
BadDiffusion & S2 & HAT & 0 & 0.00 & 50.01 & 0.2406 & 0.0004 \\
BadDiffusion & S2 & HAT & 1 & 0.00 & 49.96 & 0.2406 & 0.0004 \\
BadDiffusion & S2 & HAT & 2 & 0.00 & 50.59 & 0.2406 & 0.0004 \\
BadDiffusion & S2 & CAT & 0 & 0.00 & 50.01 & 0.3611 & 0.0002 \\
BadDiffusion & S2 & CAT & 1 & 0.00 & 49.96 & 0.3611 & 0.0002 \\
BadDiffusion & S2 & CAT & 2 & 0.00 & 50.59 & 0.3611 & 0.0002 \\
UIBDiffusion & S1 & HAT & 0 & 0.00 & 49.72 & 0.2406 & 0.0004 \\
UIBDiffusion & S1 & HAT & 1 & 0.00 & 49.05 & 0.2406 & 0.0004 \\
UIBDiffusion & S1 & HAT & 2 & 0.00 & 49.27 & 0.2406 & 0.0004 \\
UIBDiffusion & S2 & HAT & 0 & 0.00 & 46.56 & 0.2406 & 0.0004 \\
UIBDiffusion & S2 & HAT & 1 & 0.00 & 46.68 & 0.2406 & 0.0004 \\
UIBDiffusion & S2 & HAT & 2 & 0.00 & 46.71 & 0.2406 & 0.0004 \\
UIBDiffusion & S2 & CAT & 0 & 0.00 & 46.56 & 0.3611 & 0.0002 \\
UIBDiffusion & S2 & CAT & 1 & 0.00 & 46.68 & 0.3611 & 0.0002 \\
UIBDiffusion & S2 & CAT & 2 & 0.00 & 46.71 & 0.3611 & 0.0002 \\
VillanDiffusion & S1 & HAT & 0 & 0.00 & 47.10 & 0.2406 & 0.0004 \\
VillanDiffusion & S1 & HAT & 1 & 0.00 & 47.15 & 0.2406 & 0.0004 \\
VillanDiffusion & S1 & HAT & 2 & 0.00 & 47.11 & 0.2406 & 0.0004 \\
VillanDiffusion & S2 & HAT & 0 & 0.00 & 46.95 & 0.2406 & 0.0004 \\
VillanDiffusion & S2 & HAT & 1 & 0.00 & 46.72 & 0.2406 & 0.0004 \\
VillanDiffusion & S2 & HAT & 2 & 0.00 & 45.49 & 0.2406 & 0.0004 \\
VillanDiffusion & S2 & CAT & 0 & 0.00 & 46.95 & 0.3611 & 0.0002 \\
VillanDiffusion & S2 & CAT & 1 & 0.00 & 46.72 & 0.3611 & 0.0002 \\
VillanDiffusion & S2 & CAT & 2 & 0.00 & 45.49 & 0.3611 & 0.0002 \\
\bottomrule

\end{tabular}
\caption{Per-seed CIFAR-10 main results: 36 attack--scenario--target--seed rows, 1024 triggered samples per row, and 1000 inference steps.}
\label{tab:complete-main-results-cifar}
\end{table*}

\begin{table*}[t]
\centering
\small
\setlength{\tabcolsep}{3.8pt}
\begin{tabular}{lllrrrrr}
\toprule
Attack & Scen. & Target & Seed & ASR (\%) & FID & MSE & SSIM \\
\midrule
BadMerging & S1 & HAT & 0 & 0.00 & 187.87 & 0.2520 & 0.0007 \\
BadMerging & S1 & HAT & 1 & 0.00 & 177.77 & 0.2520 & 0.0007 \\
BadMerging & S1 & HAT & 2 & 0.00 & 186.74 & 0.2520 & 0.0007 \\
BadMerging & S2 & HAT & 0 & 0.00 & 141.67 & 0.2520 & 0.0007 \\
BadMerging & S2 & HAT & 1 & 0.00 & 131.29 & 0.2520 & 0.0007 \\
BadMerging & S2 & HAT & 2 & 0.00 & 138.90 & 0.2520 & 0.0007 \\
BadMerging & S2 & CAT & 0 & 0.00 & 141.67 & 0.3849 & 0.0006 \\
BadMerging & S2 & CAT & 1 & 0.00 & 131.29 & 0.3849 & 0.0006 \\
BadMerging & S2 & CAT & 2 & 0.00 & 138.90 & 0.3849 & 0.0006 \\
BadDiffusion & S1 & HAT & 0 & 0.00 & 63.79 & 0.2520 & 0.0007 \\
BadDiffusion & S1 & HAT & 1 & 0.00 & 62.02 & 0.2520 & 0.0007 \\
BadDiffusion & S1 & HAT & 2 & 0.00 & 60.39 & 0.2520 & 0.0007 \\
BadDiffusion & S2 & HAT & 0 & 0.00 & 77.64 & 0.2520 & 0.0007 \\
BadDiffusion & S2 & HAT & 1 & 0.00 & 73.99 & 0.2520 & 0.0007 \\
BadDiffusion & S2 & HAT & 2 & 0.00 & 73.37 & 0.2520 & 0.0007 \\
BadDiffusion & S2 & CAT & 0 & 0.00 & 77.64 & 0.3849 & 0.0006 \\
BadDiffusion & S2 & CAT & 1 & 0.00 & 73.99 & 0.3849 & 0.0006 \\
BadDiffusion & S2 & CAT & 2 & 0.00 & 73.37 & 0.3849 & 0.0006 \\
VillanDiffusion & S1 & HAT & 0 & 0.00 & 339.82 & 0.2520 & 0.0007 \\
VillanDiffusion & S1 & HAT & 1 & 0.00 & 338.42 & 0.2520 & 0.0007 \\
VillanDiffusion & S1 & HAT & 2 & 0.00 & 341.67 & 0.2520 & 0.0007 \\
VillanDiffusion & S2 & HAT & 0 & 0.00 & 186.65 & 0.2165 & 0.1305 \\
VillanDiffusion & S2 & HAT & 1 & 0.00 & 184.16 & 0.2179 & 0.1290 \\
VillanDiffusion & S2 & HAT & 2 & 0.00 & 187.18 & 0.2186 & 0.1266 \\
VillanDiffusion & S2 & CAT & 0 & 0.00 & 186.65 & 0.2802 & 0.1103 \\
VillanDiffusion & S2 & CAT & 1 & 0.00 & 184.16 & 0.2778 & 0.1146 \\
VillanDiffusion & S2 & CAT & 2 & 0.00 & 187.18 & 0.2801 & 0.1088 \\
\bottomrule

\end{tabular}
\caption{Per-seed CelebA-HQ main results: 27 attack--scenario--target--seed rows, 256 triggered samples per row, and 1000 inference steps.}
\label{tab:complete-main-results-celeba}
\end{table*}

\subsection{Complete Seed-0 Baseline Results}
\label{app:matched-baseline-results}

The per-case table complements main-paper
Figure~\ref{fig:matched-baseline-matrix} and
Tables~\ref{tab:cifar-s1-comparison}--\ref{tab:celeba-s2-comparison}.

\begin{table*}[t]
\centering
\small
\setlength{\tabcolsep}{1.8pt}
\begin{tabular}{lllrrrrrr}
\toprule
Dataset & Attack & Scen. & ANP & Clean FT & Diff. Soup & DMM & MaxFusion & \textbf{\method} \\
\midrule
CIFAR-10 & BadMerging & S1 & 0.0/247.5 & 0.2/53.7 & 100.0/429.4 & 99.5/62.0 & 0.0/56.0 & 0.0/50.1 \\
CIFAR-10 & BadMerging & S2 & 0.0/243.9 & 92.2/54.3 & 100.0/434.3 & 99.0/61.9 & 99.6/62.7 & 0.0/59.8 \\
CIFAR-10 & BadDiffusion & S1 & 0.0/213.8 & 0.0/56.1 & 0.0/48.7 & 0.0/51.4 & 0.0/51.9 & 0.0/51.5 \\
CIFAR-10 & BadDiffusion & S2 & 0.0/266.8 & 0.0/54.1 & 0.0/48.6 & 0.0/51.9 & 98.0/53.0 & 0.0/50.0 \\
CIFAR-10 & UIBDiffusion & S1 & 0.0/268.7 & 0.0/54.7 & 0.0/46.3 & 0.0/47.4 & 0.0/51.5 & 0.0/49.7 \\
CIFAR-10 & UIBDiffusion & S2 & 0.0/218.6 & 0.0/56.8 & 0.0/45.8 & 0.0/47.0 & 93.1/49.6 & 0.0/46.6 \\
CIFAR-10 & VillanDiffusion & S1 & 0.0/231.7 & 0.0/57.3 & 0.0/47.0 & 0.0/45.7 & 0.0/51.6 & 0.0/47.1 \\
CIFAR-10 & VillanDiffusion & S2 & 0.0/256.3 & 0.0/62.0 & 0.2/48.9 & 0.0/46.7 & 100.0/49.2 & 0.0/46.9 \\
\midrule
CelebA-HQ & BadMerging & S1 & 0.0/512.5 & 0.0/76.0 & 0.0/165.7 & 0.0/374.1 & 0.0/73.4 & 0.0/187.9 \\
CelebA-HQ & BadMerging & S2 & 0.0/453.0 & 14.1/81.0 & 48.0/141.4 & 0.0/235.1 & 100.0/154.6 & 0.0/141.7 \\
CelebA-HQ & BadDiffusion & S1 & 0.0/418.6 & 0.0/83.1 & 0.0/93.2 & 0.0/176.7 & 0.0/60.6 & 0.0/63.8 \\
CelebA-HQ & BadDiffusion & S2 & 0.0/452.7 & 0.0/140.8 & 0.0/82.2 & 0.0/205.0 & 100.0/60.9 & 0.0/77.6 \\
CelebA-HQ & VillanDiffusion & S1 & 0.0/284.9 & 0.0/81.8 & 0.0/248.4 & 0.0/117.7 & 0.0/66.9 & 0.0/339.8 \\
CelebA-HQ & VillanDiffusion & S2 & 0.0/386.4 & 0.0/102.4 & 0.0/186.0 & 0.0/138.1 & 53.5/139.6 & 0.0/186.6 \\
\bottomrule

\end{tabular}
\caption{Complete per-case seed-0 comparison. Each cell reports worst-target ASR (\%)/target-averaged FID, giving 84 method--case rows over the 14 available source cases.}
\label{tab:complete-matched-baseline}
\end{table*}

\begin{table*}[t]
\centering
\small
\setlength{\tabcolsep}{1.5pt}
\begin{tabular}{l*{10}{r}}
\toprule
& \multicolumn{2}{c}{BadMerging} & \multicolumn{2}{c}{BadDiffusion} & \multicolumn{2}{c}{UIBDiffusion} & \multicolumn{2}{c}{VillanDiffusion} & \multicolumn{2}{c}{Across attacks} \\
\cmidrule(lr){2-3}\cmidrule(lr){4-5}\cmidrule(lr){6-7}\cmidrule(lr){8-9}\cmidrule(lr){10-11}
Method & MSE$\uparrow$ & SSIM$\downarrow$ & MSE$\uparrow$ & SSIM$\downarrow$ & MSE$\uparrow$ & SSIM$\downarrow$ & MSE$\uparrow$ & SSIM$\downarrow$ & Avg. MSE & Avg. SSIM \\
\midrule
Diffusion Soup & 0.0001 & 0.9863 & 0.2406 & 0.0004 & 0.2406 & 0.0004 & 0.2406 & 0.0004 & 0.1804 & 0.2469 \\
DMM & 0.0016 & 0.9791 & 0.2406 & 0.0004 & 0.2406 & 0.0004 & 0.2406 & 0.0004 & 0.1808 & 0.2451 \\
MaxFusion & 0.2406 & 0.0004 & 0.2406 & 0.0004 & 0.2406 & 0.0004 & 0.2406 & 0.0004 & 0.2406 & 0.0004 \\
\midrule
Clean fine-tuning & 0.2083 & 0.0314 & 0.2406 & 0.0004 & 0.2406 & 0.0004 & 0.2406 & 0.0004 & 0.2325 & 0.0081 \\
ANP & 0.2397 & 0.0013 & 0.2381 & 0.0026 & 0.2412 & 0.0005 & 0.2389 & 0.0026 & 0.2394 & 0.0018 \\
\midrule
\method & 0.2406 & 0.0004 & 0.2406 & 0.0004 & 0.2406 & 0.0004 & 0.2406 & 0.0004 & 0.2406 & 0.0004 \\
\bottomrule

\end{tabular}
\caption{Matched seed-0 target MSE and SSIM for CIFAR-10 S1. These values complement main-paper Table~\ref{tab:cifar-s1-comparison} and measure similarity to the attack target.}
\label{tab:cifar-s1-similarity}
\end{table*}

\begin{table*}[t]
\centering
\small
\setlength{\tabcolsep}{1.25pt}
\begin{tabular}{ll*{10}{r}}
\toprule
& & \multicolumn{2}{c}{BadMerging} & \multicolumn{2}{c}{BadDiffusion} & \multicolumn{2}{c}{UIBDiffusion} & \multicolumn{2}{c}{VillanDiffusion} & \multicolumn{2}{c}{Across attacks} \\
\cmidrule(lr){3-4}\cmidrule(lr){5-6}\cmidrule(lr){7-8}\cmidrule(lr){9-10}\cmidrule(lr){11-12}
Method & Tgt. & MSE$\uparrow$ & SSIM$\downarrow$ & MSE$\uparrow$ & SSIM$\downarrow$ & MSE$\uparrow$ & SSIM$\downarrow$ & MSE$\uparrow$ & SSIM$\downarrow$ & Avg. MSE & Avg. SSIM \\
\midrule
Diffusion Soup & HAT & 0.0002 & 0.9822 & 0.2049 & 0.0128 & 0.2124 & 0.0272 & 0.1753 & 0.0370 & 0.1482 & 0.2648 \\
 & CAT & 0.0861 & 0.2150 & 0.3410 & 0.0092 & 0.3274 & 0.0052 & 0.2994 & 0.0250 & 0.2635 & 0.0636 \\
\addlinespace[0.1em]
DMM & HAT & 0.0012 & 0.9739 & 0.2061 & 0.0123 & 0.2166 & 0.0238 & 0.1860 & 0.0267 & 0.1525 & 0.2592 \\
 & CAT & 0.0879 & 0.2135 & 0.3427 & 0.0084 & 0.3325 & 0.0045 & 0.3168 & 0.0192 & 0.2700 & 0.0614 \\
\addlinespace[0.1em]
MaxFusion & HAT & 0.0005 & 0.9815 & 0.0042 & 0.9633 & 0.0119 & 0.9231 & 0.0015 & 0.9811 & 0.0045 & 0.9622 \\
 & CAT & 0.0903 & 0.2114 & 0.1643 & 0.1585 & 0.1051 & 0.1998 & 0.1046 & 0.2086 & 0.1161 & 0.1946 \\
\midrule
Clean fine-tuning & HAT & 0.0106 & 0.9142 & 0.2073 & 0.0117 & 0.2239 & 0.0165 & 0.1978 & 0.0181 & 0.1599 & 0.2401 \\
 & CAT & 0.1052 & 0.1964 & 0.3449 & 0.0075 & 0.3417 & 0.0021 & 0.3301 & 0.0143 & 0.2805 & 0.0551 \\
\addlinespace[0.1em]
ANP & HAT & 0.2404 & 0.0005 & 0.2396 & 0.0016 & 0.2406 & 0.0004 & 0.2383 & 0.0026 & 0.2397 & 0.0013 \\
 & CAT & 0.3601 & 0.0027 & 0.3603 & 0.0016 & 0.3616 & 0.0002 & 0.3608 & 0.0007 & 0.3607 & 0.0013 \\
\midrule
\method & HAT & 0.2406 & 0.0004 & 0.2406 & 0.0004 & 0.2406 & 0.0004 & 0.2406 & 0.0004 & 0.2406 & 0.0004 \\
 & CAT & 0.3611 & 0.0002 & 0.3611 & 0.0002 & 0.3611 & 0.0002 & 0.3611 & 0.0002 & 0.3611 & 0.0002 \\
\bottomrule

\end{tabular}
\caption{Matched seed-0 target MSE and SSIM for CIFAR-10 S2, complementing main-paper Table~\ref{tab:cifar-s2-comparison}.}
\label{tab:cifar-s2-similarity}
\end{table*}

\begin{table*}[t]
\centering
\small
\setlength{\tabcolsep}{2.0pt}
\begin{tabular}{l*{8}{r}}
\toprule
& \multicolumn{2}{c}{BadMerging} & \multicolumn{2}{c}{BadDiffusion} & \multicolumn{2}{c}{VillanDiffusion} & \multicolumn{2}{c}{Across attacks} \\
\cmidrule(lr){2-3}\cmidrule(lr){4-5}\cmidrule(lr){6-7}\cmidrule(lr){8-9}
Method & MSE$\uparrow$ & SSIM$\downarrow$ & MSE$\uparrow$ & SSIM$\downarrow$ & MSE$\uparrow$ & SSIM$\downarrow$ & Avg. MSE & Avg. SSIM \\
\midrule
Diffusion Soup & 0.2520 & 0.0007 & 0.2520 & 0.0007 & 0.2520 & 0.0007 & 0.2520 & 0.0007 \\
DMM & 0.2520 & 0.0007 & 0.2520 & 0.0007 & 0.2520 & 0.0007 & 0.2520 & 0.0007 \\
MaxFusion & 0.2520 & 0.0007 & 0.2520 & 0.0007 & 0.2520 & 0.0007 & 0.2520 & 0.0007 \\
\midrule
Clean fine-tuning & 0.2520 & 0.0007 & 0.2520 & 0.0007 & 0.2520 & 0.0007 & 0.2520 & 0.0007 \\
ANP & 0.2520 & 0.0007 & 0.2520 & 0.0007 & 0.2520 & 0.0007 & 0.2520 & 0.0007 \\
\midrule
\method & 0.2520 & 0.0007 & 0.2520 & 0.0007 & 0.2520 & 0.0007 & 0.2520 & 0.0007 \\
\bottomrule

\end{tabular}
\caption{Matched seed-0 target MSE and SSIM for CelebA-HQ S1, complementing main-paper Table~\ref{tab:celeba-s1-comparison}.}
\label{tab:celeba-s1-similarity}
\end{table*}

\begin{table*}[t]
\centering
\small
\setlength{\tabcolsep}{1.75pt}
\begin{tabular}{ll*{8}{r}}
\toprule
& & \multicolumn{2}{c}{BadMerging} & \multicolumn{2}{c}{BadDiffusion} & \multicolumn{2}{c}{VillanDiffusion} & \multicolumn{2}{c}{Across attacks} \\
\cmidrule(lr){3-4}\cmidrule(lr){5-6}\cmidrule(lr){7-8}\cmidrule(lr){9-10}
Method & Tgt. & MSE$\uparrow$ & SSIM$\downarrow$ & MSE$\uparrow$ & SSIM$\downarrow$ & MSE$\uparrow$ & SSIM$\downarrow$ & Avg. MSE & Avg. SSIM \\
\midrule
Diffusion Soup & HAT & 0.0417 & 0.7188 & 0.2516 & 0.0009 & 0.1826 & 0.1964 & 0.1586 & 0.3054 \\
 & CAT & 0.1245 & 0.5341 & 0.3844 & 0.0008 & 0.2865 & 0.1627 & 0.2651 & 0.2325 \\
\addlinespace[0.1em]
DMM & HAT & 0.2251 & 0.0431 & 0.2520 & 0.0007 & 0.1952 & 0.1637 & 0.2241 & 0.0692 \\
 & CAT & 0.3531 & 0.0354 & 0.3849 & 0.0006 & 0.3084 & 0.1365 & 0.3488 & 0.0575 \\
\addlinespace[0.1em]
MaxFusion & HAT & 0.0092 & 0.8439 & 0.0690 & 0.5733 & 0.1125 & 0.4571 & 0.0636 & 0.6248 \\
 & CAT & 0.1118 & 0.5460 & 0.0188 & 0.8139 & 0.0790 & 0.6192 & 0.0699 & 0.6597 \\
\midrule
Clean fine-tuning & HAT & 0.1197 & 0.2874 & 0.2519 & 0.0008 & 0.2167 & 0.0757 & 0.1961 & 0.1213 \\
 & CAT & 0.3509 & 0.0131 & 0.3847 & 0.0007 & 0.3444 & 0.0515 & 0.3600 & 0.0218 \\
\addlinespace[0.1em]
ANP & HAT & 0.0744 & 0.3353 & 0.2520 & 0.0007 & 0.2513 & 0.0008 & 0.1926 & 0.1123 \\
 & CAT & 0.3844 & 0.0006 & 0.3849 & 0.0006 & 0.3806 & 0.0007 & 0.3833 & 0.0006 \\
\midrule
\method & HAT & 0.2520 & 0.0007 & 0.2520 & 0.0007 & 0.2165 & 0.1305 & 0.2402 & 0.0440 \\
 & CAT & 0.3849 & 0.0006 & 0.3849 & 0.0006 & 0.2802 & 0.1103 & 0.3500 & 0.0372 \\
\bottomrule

\end{tabular}
\caption{Matched seed-0 target MSE and SSIM for CelebA-HQ S2, complementing main-paper Table~\ref{tab:celeba-s2-comparison}.}
\label{tab:celeba-s2-similarity}
\end{table*}

\section{Additional Analysis}
\label{app:additional-analysis}

\subsection{R3S Diagnostic}
\label{app:r3s-diagnostic}

\begin{figure}[t]
\centering
\includegraphics[width=\columnwidth]{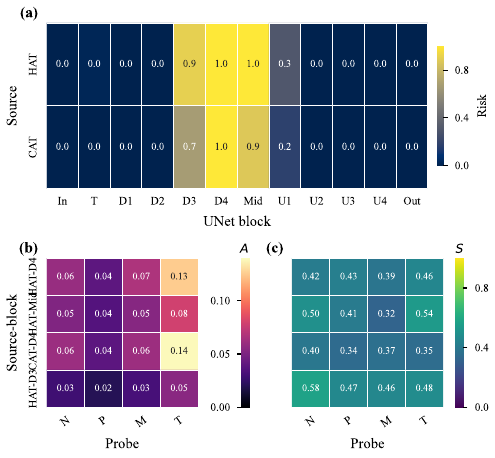}
\caption{R3S diagnostic on held-out VillanDiffusion CIFAR-10 S2 (Diffusion Soup HAT ASR 0.20\%), used for mechanism inspection rather than as an attack-frequency estimate: (a) source--block risk; (b,c) amplification and path stability for the top four pairs. N/P/M/T denote the probe families.}
\label{fig:r3s-diagnostic-appendix}
\end{figure}

The diagnostic places the largest risk in D3, D4, and the middle block
of both sources. The top pairs vary across probes and ratio paths.
Because R3S also uses clean utility and hierarchical aggregation, this
figure explains score composition rather than causal localization or
attack prevalence.

\subsection{Complete Component Ablation}
\label{app:component-ablation}

\begin{table*}[t]
\centering
\small
\setlength{\tabcolsep}{7.0pt}
\begin{tabular}{lrr}
\toprule
Configuration & Worst ASR (\%) & FID \\
\midrule
Full \method & 0.000 & 57.11 \\
w/o ratio stability & 0.000 & 49.67 \\
w/o clean utility & 0.000 & 48.51 \\
w/o clean backoff & 0.000 & 56.48 \\
w/o source floor & 97.949 & 406.70 \\
Renormalized source weights & 100.000 & 438.31 \\
w/o output-time evidence & 4.655 & 68.46 \\
Source-only attenuation & 0.000 & 58.53 \\
w/o L1 bound & 0.000 & 55.21 \\
Calibration $n=4$ & 0.000 & 47.88 \\
Calibration $n=16$ & 100.000 & 130.88 \\
Noise + patch probes & 0.000 & 51.62 \\
Mask + texture probes & 100.000 & 268.00 \\
\bottomrule
\end{tabular}
\caption{Complete 13-configuration component study on BadMerging CIFAR-10 S2. FID is computed from no-trigger samples.}
\label{tab:complete-component-ablation}
\end{table*}

Table~\ref{tab:complete-component-ablation} reports target-matching ASR
and clean FID. Removing the source floor, renormalizing source weights,
using 16 calibration images, or using only mask and texture probes
restores high HAT ASR.

\subsection{Controls and Sensitivity}
\label{app:sensitivity-controls}

\begin{table*}[t]
\centering
\small
\setlength{\tabcolsep}{7.0pt}
\begin{tabular}{lrr}
\toprule
Configuration & Worst ASR (\%) & FID \\
\midrule
Reference only & 0.000 & 52.35 \\
Scalar global attenuation & 0.000 & 51.73 \\
Matched random blocks & 0.000 & 56.04 \\
\midrule
Strength $\eta=0$ & 100.000 & 435.41 \\
Strength $\eta=0.25$ & 100.000 & 407.17 \\
Strength $\eta=0.50$ & 100.000 & 252.97 \\
Strength $\eta=0.75$ & 0.749 & 93.05 \\
\midrule
Clean--clean reference & 0.000 & 50.92 \\
Clean--clean soup & 0.000 & 50.36 \\
Clean--clean \method & 0.000 & 51.21 \\
\midrule
Random blocks (seed 2028) & 0.000 & 94.60 \\
Random blocks (seed 2029) & 0.000 & 82.00 \\
\bottomrule
\end{tabular}
\caption{Focused controls, attenuation sensitivity, clean--clean checks, and two additional matched-random seeds. All rows use three seeds, 1024 samples per seed, and 1000 inference steps.}
\label{tab:auxiliary-analysis}
\end{table*}

The clean--clean rows compare the trusted reference, the intended
two-clean-source soup, and \method. Their FIDs are 50.92, 50.36, and
51.21, and every clean output passes the validity check. The
matched-random result also varies across block-selection seeds: all
three choices retain zero ASR, but their clean FID is unstable. This
focused control does not establish R3S as the only route to zero ASR.

\subsection{R3S-Aware Adaptive Attack}
\label{app:adaptive-attack}

The adaptive comparison changes the attack training objective but
keeps the \method configuration fixed.

\begin{table}[t]
\centering
\small
\setlength{\tabcolsep}{5.0pt}
\begin{tabular}{lrr}
\toprule
Attack setting & ASR (\%) & FID \\
\midrule
Standard BadMerging & $0.0\pm0.0$ & $49.58\pm0.54$ \\
R3S-aware BadMerging & $0.0\pm0.0$ & $52.47\pm0.47$ \\
\bottomrule
\end{tabular}
\caption{CIFAR-10 S1 target-matching stress test over three seeds. The attacker optimizes against the exact R3S score; \method is unchanged.}
\label{tab:adaptive-attack}
\end{table}

\section{Efficiency and Qualitative Results}

The clean-only audit covers 25 BadMerging CIFAR-10 S2 configurations.
Every clean row has 100\% valid outputs under standard no-trigger
sampling.

\begin{figure*}[t]
\centering
\includegraphics[width=0.96\textwidth]{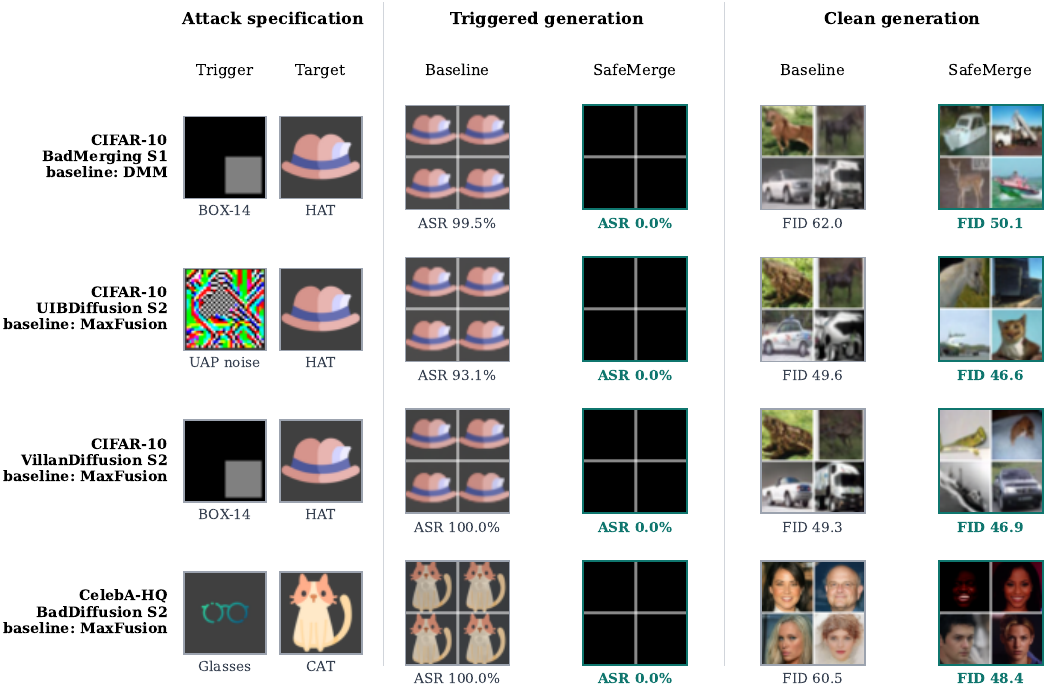}
\caption{Fixed seed-0 samples from four attack settings. Triggered samples illustrate target matching, while clean samples provide a visual counterpart to FID. The reported metrics use all evaluation samples.}
\label{fig:appendix-qualitative}
\end{figure*}

Method cost is the sum of the risk diagnostic, five candidate builds,
and the clean guard. Across 14 source cases, it is
$957.9\pm1220.7$ GPU-seconds per case; the maximum allocated memory is
3005 MiB. Shared calibration export and the final checkpoint copy are
excluded. The selected checkpoint has the same inference structure as
an ordinary merged DDPM.

\end{document}